\documentclass[preprint,showpacs,preprintnumbers]{revtex4}
\usepackage{epsfig}
\begin{document}
\title{SYMMETRY AND ELECTRONIC STRUCTURE OF NOBLE METAL NANOPARTICLES AND THE ROLE OF RELATIVITY}
\author{Hannu H\"{a}kkinen}
\affiliation{Department of Physics, NanoScience Center, FIN-40014
University of Jyv\"{a}skyl\"{a}, Finland}
\author{Michael Moseler} \affiliation{Fraunhofer Institute for mechanics
of materials, Wohlerstra{\ss}e 11, D-79108 Freiburg, Germany}
\author{ Oleg Kostko, Nina Morgner, Margarita Astruc Hoffmann, and
Bernd v. Issendorff}\email{bernd.von.issendorff@uni-freiburg.de}
\affiliation{Fakult\"{a}t f\"{u}r Physik, Universit\"{a}t Freiburg,
Stefan-Meier-Stra{\ss}e 21, D-79104 Freiburg, Germany}

\date{\today}
\begin{abstract}
High resolution photoelectron spectra of cold mass selected
Cu$_n^-$, Ag$_n^-$ and Au$_n^-$ with n =53-58 have been measured
at a photon energy of 6.42 eV. The observed electron density of
states is not the expected simple electron shell structure, but
seems to be strongly influenced by electron-lattice interactions.
Only Cu$_{55}^-$ and Ag$_{55}^-$ exhibit highly degenerate states.
This is a direct consequence of their icosahedral symmetry, as is
confirmed by density functional theory calculations. Neighboring
sizes exhibit perturbed electronic structures, as they are formed
by removal or addition of atoms to the icosahedron and therefore
have lower symmetries. Gold clusters in the same size range show
completely different spectra with almost no degeneracy, which
indicates that they have structures of much lower symmetry. This
behaviour is related to strong relativistic bonding effects in
gold, as demonstrated by ab initio calculations for Au$_{55}^-$.

\end{abstract}
\maketitle

Understanding the energetically most favorable structures that
aggregates of metal atoms inherently adopt during their formation
process is one of the long-standing issues in the science of
clusters and nanoparticles\cite{deheerrev,tpmartin-shell}. The
atomic structure and its symmetry are intimately related to the
electronic structure, which in turn defines the electrical,
optical and chemical properties of the particle. Resolving the
atomic structures of nanoclusters therefore represents an
important preliminary for their controlled use in future
nanotechnologies. Noble metal clusters and nanoparticles have
attracted much attention recently, as they are considered
promising candidates for applications e.g. in catalysis
\cite{kata,kata-au}, labeling \cite{au-label} or
photonics\cite{nanoopt}. Two size ranges of these particles have
been characterized in detail: small clusters with up to 21 atoms
have been studied by high resolution photoelectron and optical
spectroscopy as well as a number of theoretical
studies\cite{gantecu,ganteag,meiwes,koutecky99,landman-cu7-au7,aumob,hannu-wang-au,wang-Au-20}.
The results show that in this still rather ``molecular" size range
the clusters tend to adopt a variety of low symmetry structures
(the tetrahedral Au$_{20}^-$ being a prominent
exception\cite{wang-Au-20}). Much larger particles, with hundreds
or thousands of atoms, have been studied by electron microscopy
and electron diffraction\cite{ag-diffr,marks}. It turned out that
these particles adopt well-ordered structures with surprisingly
diverse overall symmetries. Icosahedral, decahedral and
(bulk-like) octahedral forms are often observed at the same time,
with the preferred structural motif depending sensitively on the
particle formation process. Experimental structural information
for medium sized (around 50 atoms and up) Cu, Ag and Au clusters
was scarce up to now. Measurements of ionization
potentials\cite{kappes-ag} and UV-photoelectron spectroscopy
studies\cite{cheshcuau} gave evidence for the existence of some
free electron shell structure, which lead to the general
perception that the properties of noble metal clusters in this
size range can be fully explained by the free electron model.

We have performed high resolution UV-photoelectron spectroscopy on
free, cold, size selected noble metal clusters. The setup is the
same as used in earlier experiments\cite{al32k}. Copper, silver
and gold clusters were produced by magnetron discharge sputtering
of a metal target inside a liquid nitrogen cooled aggregation
tube, through which a mixture of helium and argon is flowing at a
pressure of about 0.5 mbar. As the discharge produces many charged
condensation seeds, a large portion of the clusters formed in the
cold gas flow is negatively (or positively) charged. After having
covered a distance of 25 cm the gas with the clusters inside
expands into the vacuum through an adjustable aperture (typically
4 mm diameter). The clusters pass a skimmer and are inserted into
a double-reflectron time-of-flight mass spectrometer, which is
used to select a single cluster size. These clusters are
decelerated and inserted into a magnetic bottle time-of-flight
photoelectron spectrometer, where they are irradiated by photons
from an ArF-excimer laser (hv=6.42 eV). The flight time
distribution of the emitted electrons is measured and converted
into a binding energy distribution. The electron spectrometer has
an energy resolution of about E/dE = 40. It has been calibrated by
measuring the known spectra of the monomer anions, which leads to
an error of the energy axis of less than 30 meV. Typically, the
photoelectron spectra have been averaged over 30000 laser shots at
a repetition rate of 100 Hz. We estimate the temperature of the
clusters to be $200\pm 50$ K.

\begin{figure}[t]
\begin{center}
\epsfig{file=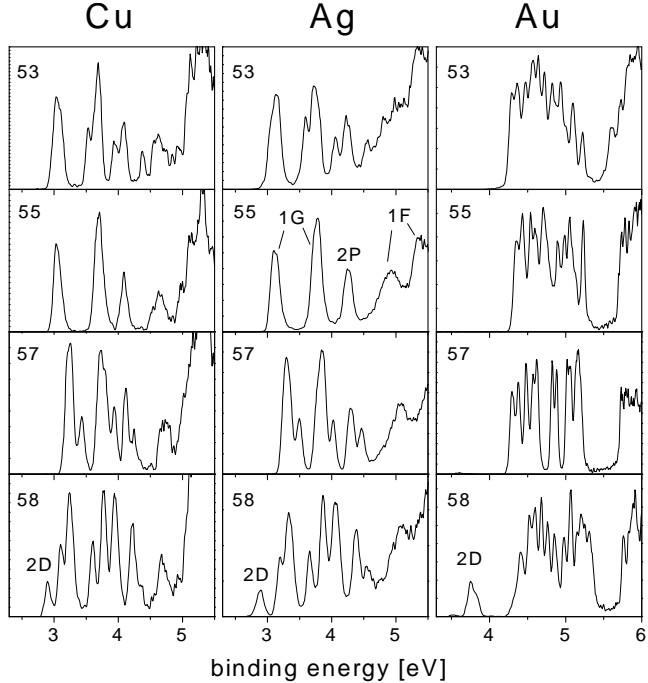,clip=,width=8.5cm}
\caption{Photoelectron
spectra of Cu$_N^-$, Ag$_N^-$ and Au$_N^-$ (N=53,55,57,58)
obtained at a photon energy of 6.42 eV.}
\end{center}
\end{figure}

The spectra obtained for Cu$_n^-$, Ag$_n^-$ and Au$_n^-$ with
n=53-58 are shown in Fig. 1. In principle these are direct images
of the electronic density of states. In the bulk the electronic
structure of nobel metals is characterized by a half-filled and
rather free-electron like band formed from the atomic s-orbitals,
intersected some eV below the Fermi energy by the so-called
d-band, which is formed from the rather localized atomic
d-orbitals. As has been discussed in detail already by Taylor et
al.\cite{cheshcuau} this structure is clearly visible in the
photoelectron spectra: the uppermost part of the distributions can
be identified as being dominantly s-electron derived, while the
onset of the d-bands can be observed roughly 2-3 eV  below the
uppermost state. The s-band part of the electronic density of
states should therefore exhibit the same discrete structure as
observed e.g. for alkali clusters\cite{deheerrev,napos}. The
57-atom cluster anions contain 58 valence s-electrons, and in the
spherical droplet approximation should have the closed shell
configuration 1S$^2$1P$^6$1D$^{10}$2S$^2$1F$^{14}$2P$^6$1G$^{18}$
(for clarity we refer to free electron angular momentum states
with a capital letter). Adding one atom should lead to the
appearance of a singly occupied 2D orbital. Indeed for all three
metals a new peak appears for size 58, which indicates the
formation of a new shell. However, the free electron model also
predicts that in the section of the s-band visible here only four
shells should be present: 1F, 2P, 1G and 2D. The structure of  the
photoelectron spectra obviously is more complicated than that,
which hints at a relatively strong perturbation of the shell
structure by electron-lattice interaction. The only clusters
exhibiting a clear structure of highly degenerate states are
Cu$_{55}^-$ and Ag$_{55}^-$. Such degeneracies can only be
produced by a highly symmetric atomic structure. Au$_{55}^-$, on
the other hand, exhibits a complex spectrum without significant
level bunching; it therefore seems not to adopt any of the atomic
symmetries that lead to pronounced degeneracy of electron shells.
This is in accordance with earlier mass spectroscopic measurements
which demonstrated that 55 is a strong atomic magic number for
silver, but not for gold\cite{lutz-ag,lutz-au}.

We will now discuss these findings in the light of our theoretical
results. The atomic and electronic structures of silver and gold
clusters were calculated  by density functional theory in
combination with Born-Oppenheimer molecular
dynamics\cite{BO-LSD-MD}, including self-consistent gradient
corrections\cite{gga}. The interaction of the d$^{10}$s$^1$
valence electrons of Ag and Au with the atomic cores was described
by scalar-relativistic norm-conserving
pseudopotentials\cite{pseudopot}. For Au a non-relativistic
pseudopotential was used as well, in order to check the influence
of relativity. The Kohn-Sham orbitals were expanded in a plane
wave basis with a kinetic energy cutoff of 62 Ry. The method does
not employ a supercell, i.e., a periodic image of the atoms, and
is thus suitable for accurate calculations for charged finite
clusters\cite{BO-LSD-MD}.
 As a full simulated-annealing scheme with ab-initio determination
 of the electronic structure at each dynamic step still is
computationally too expensive for this size of d-metal clusters
(having a total number of valence electrons of about 600), we
adopted an alternative strategy. We made use of the already
existing knowledge about possible structures of nanosized metal
clusters, which has been obtained by simulated annealing and
genetic algorithm optimization of clusters described by classical
pairwise and many-atom interatomic potentials. An extensive
collection of such structures, known as the Cambridge Cluster
Database (CCD), is publicly available on the World Wide
Web\cite{ccd}. For the structural optimization of Ag$_{55}^-$ and
Au$_{55}^-$ we chose 6 candidates representing different
structural motifs: closed atomic shell icosahedral (ICO),
decahedral (DECA), and cuboctahedral (CUBO) structures, and the
optimum structures found for classical Sutton-Chen (SC), glue, and
Morse potentials. These geometries were then optimized by ab
initio molecular dynamics simulations, allowing the clusters to
relax for 1~ps under the influence of a global friction force. In
none of the cases this relaxation lead to a severe change of the
overall structure.

\begin{figure}
\begin{center}
\epsfig{file=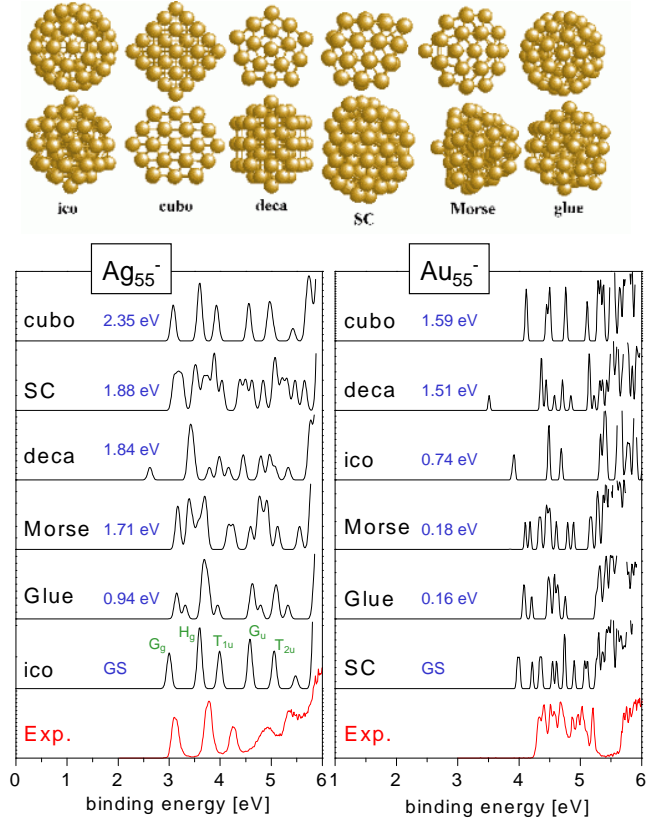,clip=,width=8.5cm} \caption{Structures and
density of states (DOS) of Ag and Au clusters obtained via density
functional calculations. (a) Six candidate structures (displayed
are the fully relaxed optimal structures for gold) for Ag$_{55}^-$
and Au$_{55}^-$ representing different structural motifs: closed
atomic shell icosahedral (ICO), decahedral (DECA), and
cuboctahedral (CUBO) structures, and clusters optimized previously
by classical Sutton-Chen (SC), Glue, and short ranged Morse
potentials (b) DOS of the six structures (black curves) compared
to the experimental photoelectron spectra (red curves) for Ag
(left panel) and Au (right panel). The numbers denote the energy
difference to the most stable structure (GS).}
\end{center}
\end{figure}

The optimized structures and their electronic densities of states
(DOS) as well as the measured spectra are shown in Fig. 2. One can
see that only two of them, the icosahedron\cite{note} and the
cuboctahedron, show a clear shell structure like the one observed
in the experiment. All other isomers exhibit much less distinct
level bunching. Of the six structures considered, the icosahedral
cluster ICO is the clear ground state. Its calculated DOS matches
very well the experimental photoelectron spectrum. This allows us
to identify the shells visible in the measured spectrum as the
icosahedral orbitals T$_{2u}^6$G$_u^8$T$_{1u}^6$H$_g^{10}$G$_g^6$.
These levels can be seen as being produced from the angular
momentum eigenstates of the spherical droplet by crystal field
splitting. The lattice of the cluster represents a perturbing
potential of icosahedral symmetry. Group theory tells
us\cite{Berry-Al} that such a perturbation does not lift the
degeneracy of the 2P shell, but splits the 1F and 1G shell into
two subshells each, as indicated in Fig. 1. The separation of
these subshells can therefore be taken as a measure for the
strength of the electron lattice interaction. The cuboctahedral
cluster, though exhibiting a DOS very similar to that of the ICO
cluster, can safely be excluded to be present in the experiment
because of its significantly higher energy. All other isomers will
have a more complex DOS. So although we cannot exclude that other
low symmetry isomers exist which are energetically closer to the
ICO than the ones presented here, the purity of the experimental
spectrum demonstrates that the icosahedron is by far the dominant
structure for the Cu$_{55}^-$ and Ag$_{55}^-$ clusters produced by
our source.

As the atomic symmetry of these two clusters leads to the high
degeneracy in their electronic structure\cite{note}, it is
interesting to see how it will change if the symmetry is
perturbed. In Fig. 1 one can observe that if the cluster size is
increased or decreased by two atoms, the three upper peaks in the
spectra split up. This splitting can be most easily understood for
the case of the T$_{1u}$ (2P) orbital. Assuming that the
geometries of the clusters are based on the 55-atom icosahedron,
Ag$_{53}^-$ has two vacancies in the outermost atom shell and
therefore a slightly oblate shape, whence Ag$_{57}^-$ has two
adatoms and is prolate. In the case of a P-type orbital this will
lead to a splitting into the P$_x^2$P$_y^2$P$_z^2$ subshells, with
1:2 and 2:1 intensity ratios of the lower:higher binding energies
for oblate and prolate shapes, respectively. This is exactly what
one can observe in Fig. 1. The same can be seen in the theoretical
spectra of Ag$_{57}^-$ in Fig. 3, where the calculated DOS of the
three lowest energy isomers is shown. Interestingly the splitting
of the shells is more pronounced in the calculation than in the
experiment. This could be a temperature effect:  in the
calculation the geometries are fully relaxed at 0 K; here the
attached atoms produce some distortion of the icosahedral core and
thereby some additional perturbation of the shell structure. In
the experiment the finite cluster temperature (about 150-250 K)
will probably lead to a larger average bond length of the surface
atoms and therefore to a reduced effect on the icosahedral core.
Additionally the attached atoms will probably be more mobile than
the energy differences between the different isomers indicate. In
order to check this we show a simple average of the three
calculated spectra as well. The better agreement of this averaged
spectrum with the measured one indicates that indeed more than one
isomer is present in the experiment, which is probably due to a
constant movement of the attached atoms on the icosahedral core.

 Let us now turn to the case of gold. The experimental spectrum of the Au$_{55}^-$ cluster
 is drastically different from those of Ag$_{55}^-$
and Cu$_{55}^-$ (Fig. 1). The visible part of the ``s-band" is
highly structured , but does not show any trace of a shell
pattern. This can be taken as direct evidence that a free
Au$_{55}^-$ does not adopt icosahedral or cuboctahedral symmetry.
Indeed from the computations we find several low-symmetry
structures (SC, GL, MO) for Au$_{55}^-$ below the ICO. The
calculated DOS of these isomers are in a qualitative agreement
with the measured spectra, exhibiting no significant level
bunching. Perfect agreement cannot be expected, as spin-orbit
coupling, which is not taken  into account in the calculation,
will further perturb the DOS. Additionally, one cannot assume that
the SC isomer is the true ground state of Au$_{55}^-$, given the
vast amount of possible low-symmetry structures (in a previous
calculation using semiempirical potentials 360 disordered isomers
have been identified that were energetically favorable to the ICO
\cite{soler}). Nevertheless both our experiment and our
calculation give strong evidence for a low symmetry ground state
of Au$_{55}^-$.
\begin{figure}
\begin{center}
\epsfig{file=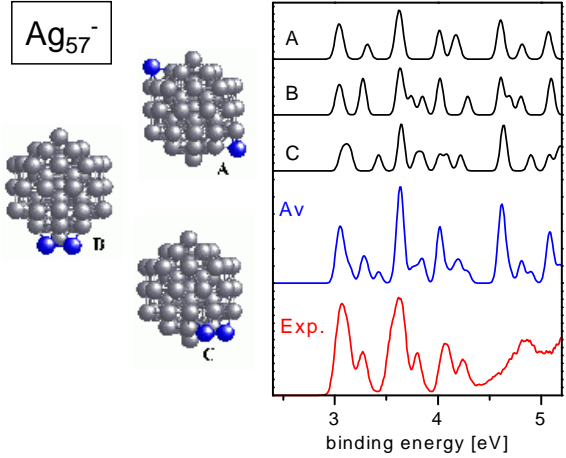,clip=,width=7.5cm} \caption{Three
icosahedral-based candidate structures (A,B,C) of Ag$_{57}^-$ and
their calculated DOS as well as an average of A,B and C (AV) and
the experimental photoelectron spectrum (Exp.). The calculated
energies of the isomers A and B with respect to the ground state
isomer C are 0.63 eV and 0.23 eV, respectively.}
\end{center}
\end{figure}
So why are gold clusters so different from silver or copper
clusters? The answer is found by comparing the results of a
nonrelativistic calculation with those of a scalar-relativistic
one. It turns out that a fictitious, nonrelativistic Au$_{55}^-$
behaves very similar to Cu$_{55}^-$ or Ag$_{55}^-$, having a clear
ICO ground state and an almost identical DOS. Only in the
scalar-relativistic calculation the lower symmetry isomers are
preferred, which is due to a change in the nature of the
interatomic bonding. The Au atom is known to be ``the most
relativistic element below Fermium"\cite{pyy}, displaying a strong
outer shell (6s) contraction and a reduced 5d--6s energy gap. This
leads to significant s-d hybridization and direct d-d bonding
effects, which for instance are also responsible for the
anomalous, planar ground-state structures of anionic Au$_N^-$
clusters up to about $N=12$
\cite{landman-cu7-au7,aumob,hannu-wang-au}. The most obvious
consequences of these relativistic effects are a shortening of the
interatomic bond lengths and a drastic increase of the bulk
modulus. In a calculation of the Au$_2$ potential energy curve we
have obtained an increase of the curvature from 5.3 eV/\AA$^2$ to
11.6 eV/\AA$^2$ when including relativistic effects. The same
trend was reported earlier for the calculated bulk modulus of
crystalline Au, where a relativistic treatment increased the value
from 108 GPa to 182 GPa\cite{aubulk-rel}. A recent general study
of metal cluster structures demonstrated that it is mainly this
high bulk modulus which leads to the different behaviour of gold
and silver clusters\cite{soler}.

The fact that Au$_{55}^-$ prefers a low-symmetry structure can
therefore be clearly identified as a relativistic effect. This
demonstrates the important influence of relativity on the
structure of mesoscopic heavy atom particles.

This work was supported by the Academy of Finland
and the Deutsche Forschungsgemeinschaft. The experimental work was
done by O. Kostko, N. Morgner, M. Astruc Hoffmann and B. v.
Issendorff. Computations were performed by H.H\"{a}kkinen at the
CSC--Scientific Computing Ltd. in Espoo, Finland, and by
M.Moseler at HLRS in Stuttgart and NIC in J\"{u}lich.


\begin{thebibliography}{}
\bibitem{note} We call these structures icosahedral although the
partial occupation of the uppermost G$_g$ shell leads to a
Jahn-Teller (JT) deformation; due to the rigidity of the lattice
this deformation is, however, very small. In our unconstrained MD
optimization we find JT-induced changes of the bondlengths of less
than 2\%; this results in a total width e.g. of the G$_g$ derived
band of only 0.04~eV.
\end{thebibliography}
\end{document}